\documentclass[12pt]{article}

\renewcommand{\d}{{\rm d}}

\newcommand{\son}{\sigma^1}
\newcommand{\tson}{\tilde{\sigma}^1}

\begin{document}

\title{On the Faddeev gravity on the piecewise flat manifold}
\author{V.M. Khatsymovsky \\
 {\em Budker Institute of Nuclear Physics} \\ {\em of Siberian Branch Russian Academy of Sciences} \\ {\em
 Novosibirsk,
 630090,
 Russia}
\\ {\em E-mail address: khatsym@inp.nsk.su}}
\date{}
\maketitle
\begin{abstract}
We study the Faddeev formulation of gravity in which the metric is composed of vector fields. This system is reducible with the help of the equations of motion to the general relativity. The Faddeev action is evaluated for the piecewise flat ansatz for these fields when the metric corresponds to the flat interior of the 4-simplices of the general simplicial complex. Thereby an analogue of the Regge action in the usual general relativity is obtained. A peculiar feature of the Faddeev gravity is finiteness of the action on the discontinuous fields, and this means possibility of the complete independence of the fields in the different 4-simplices or incoincidence of the 4-simplices on their common faces. The earlier introduced analogue of the Barbero-Immirzi parameter for the Faddeev gravity is taken into account. There is some freedom in defining the Faddeev action on the piecewise flat manifold, and the task is set to make use of this freedom to ensure that this discrete system be reducible with the help of the {\it discrete} equations of motion to the analogous discrete general relativity (Regge calculus).
\end{abstract}

keywords: general relativity; piecewise flat spacetime


MSC classes: 83C27

\section{ Introduction}

Recently Faddeev has proposed \cite{Fad} a new formulation of the Einstein's gravity described by a set of ten covariant vector fields $f^A_\lambda (x)$. Here, the Latin capitals $A, B, \dots = 1, \dots , 10$ refer to an Euclidean (or Minkowsky) ten-dimensional spacetime, and the Greek indices $\lambda, \mu, \dots = 1, 2, 3, 4$ refer to our four-dimensional spacetime. To simplify notations, we consider the case of the Euclidean metric signature for both the spaces. Our usual metric is a composite field, $g_{\lambda \mu} = f^A_\lambda f_{\mu A}$, and thus $f^A_\lambda$ is a ten-dimensional tetrad. An important point of this approach is introducing a connection
\begin{equation}                                                            
\tilde{\Gamma}_{\lambda \mu\nu} = f^A_\lambda f_{\mu A, \nu} ~~~ (f_{\mu A, \nu} \equiv \partial_\nu f_{\mu A}), ~~~ \tilde{\Gamma}^\lambda_{\mu\nu} = g^{\lambda\rho} \tilde{\Gamma}_{\rho \mu \nu}
\end{equation}

\noindent alternative to the unique torsion-free Levi-Civita one, $\Gamma^\lambda_{ \mu\nu}$. Then the curvature tensor reads
\begin{equation}                                                            
\hspace{-15mm} S^\lambda_{\mu \nu \rho} = \tilde{\Gamma}^\lambda_{\mu\rho, \nu} - \tilde{\Gamma}^\lambda_{\mu\nu, \rho} + \tilde{\Gamma}^\lambda_{\sigma\nu} \tilde{\Gamma}^\sigma_{\mu\rho} - \tilde{\Gamma}^\lambda_{\sigma\rho} \tilde{\Gamma}^\sigma_{\mu\nu}
= \Pi^{AB} (f^\lambda_{A, \nu} f_{\mu B, \rho} - f^\lambda_{A, \rho} f_{\mu B, \nu}).
\end{equation}

\noindent Here, $\Pi_{AB} = \delta_{AB} - f^\lambda_A f_{\lambda B}$ is the {\it projector onto the vertical directions}, or onto the six-dimensional subspace in the ten-dimensional spacetime orthogonal to that one spanned by the tetrad. Note that this projector makes the usual derivatives equivalent to the covariant ones,
\begin{equation}                                                            
\Pi^{AB} f_{\lambda B, \mu} = \Pi^{AB} \nabla_\mu f_{\lambda B}.
\end{equation}

\noindent The action takes the form
\begin{equation}\label{Fad action}                                          
\hspace{-10mm} S = \int {\cal L} \d^4 x = \int S^\lambda_{\mu \lambda \rho} g^{\mu \rho} \sqrt {g} \d^4 x = \int \Pi^{AB} (f^\lambda_{A, \lambda} f^\mu_{B, \mu} - f^\lambda_{A, \mu} f^\mu_{B, \lambda}) \sqrt {g} \d^4 x.
\end{equation}

Applying the operator $\Pi_{AB} \delta / \delta f^\lambda_B$ to this action, we get the {\it vertical components} of the equations of motion,
\begin{equation}\label{V lambda A}                                          
b^\mu{}_{\mu A} T^\nu_{\lambda \nu} + b^\mu{}_{\lambda A} T^\nu_{\nu \mu} + b^\mu{}_{\nu A} T^\nu_{\mu \lambda} = 0.
\end{equation}

\noindent Here, $b^A_{\lambda \mu} = \Pi^{AB} f_{\lambda B, \mu} \equiv \Pi^{AB} \nabla_\mu f_{\lambda B}$, and $T^\lambda_{ \mu \nu } = f^{\lambda A} (f_{\mu A , \nu} - f_{\nu A , \mu})$ is torsion. Since the index $A$ in the system of equations (\ref{V lambda A}) is that of some projected by $\Pi_{AB}$ expression, it takes on effectively 6 values. Thus we have $4 \times 6 = 24$ independent equations. If these equations are considered as a linear system for $T^\lambda_{\mu \nu}$, the number of these equations might be sufficient to ensure that $4 \times 6 = 24$ components of $T^\lambda_{\mu \nu}$ be zero. Namely, the matrix of the system (\ref{V lambda A}) is a square $24 \times 24$ matrix (a function of $b^\lambda{}_{\mu A}$), and the determinant of it is nonzero for random values of $b^\lambda{}_{\mu A}$ \cite{Fad}. Therefore, it is assumed that the vertical equations of motion are equivalent to the vanishing of torsion: $T^\lambda_{\mu \nu} = 0$. This means that $\tilde{\Gamma}^\lambda_{\mu\nu} = \Gamma^\lambda_{\mu\nu}$. Substituting this back into $S$, we obtain the Hilbert-Einstein action. Therefore, the other components of the equations of motion are the Einstein equations.

We have found \cite{our} that the Faddeev action can be modified by adding to it some parity odd term,
\begin{eqnarray}\label{Fad action+Imm term}                               
\int {\cal L} \d^4 x & = & \int \Pi^{AB} \left [ (f^\lambda_{A, \lambda} f^\mu_{B, \mu} - f^\lambda_{A, \mu} f^\mu_{B, \lambda}) \sqrt {g} - \frac{1}{\gamma_F} \epsilon^{\lambda \mu \nu \rho} f_{\lambda A, \mu} f_{\mu B, \rho} \right ] \d^4 x \\ \label{Fad action+Imm term 2} & = & \int g^{\lambda \nu} g^{\mu \rho} S_{\lambda \mu \nu \rho} \sqrt {g} \d^4 x + \frac{1}{2 \gamma_F} \int \epsilon^{\lambda \mu \nu \rho} S_{\lambda \mu \nu \rho} \d^4 x.
\end{eqnarray}

\noindent Here $\gamma_F$ is an analog of the known Barbero-Immirzi parameter $\gamma$ \cite{Barb,Imm} because it arises in the aspect of the first order formalism for the Faddeev gravity considered in \cite{our} upon adding certain term to the action analogous to that one parameterized by $\gamma$ and added to the Cartan-Weyl form of the Hilbert-Einstein action \cite{Holst,Fat}. Contrary to the case of $\gamma$, the term parameterized by $\gamma_F$ in the first order formalism for the Faddeev gravity does not vanish on the equations of motion for the connection. This leads to the above modification of the genuine Faddeev action. However, this term disappears at the second stage upon partial using the equations of motion like (\ref{V lambda A}). Indeed, modification of (\ref{V lambda A}) for the action (\ref{Fad action+Imm term}) looks as
\begin{equation}\label{V lambda A-modified}                                 
\hspace{-10mm} b^\mu{}_{\mu A} T^\nu_{\lambda \nu} + b^\mu{}_{\lambda A} T^\nu_{\nu \mu} + b^\mu{}_{\nu A} T^\nu_{\mu \lambda} + \frac{\epsilon^{\tau \mu \nu \rho}}{2 \gamma_F \sqrt{g}} (g_{\lambda \sigma} g_{\kappa \tau} - g_{\lambda \tau} g_{\kappa \sigma}) b^{\kappa}{}_{\rho A} T^\sigma_{\mu \nu} = 0.
\end{equation}

\noindent  The modification made seems to be not crucial for that the determinant of this linear system be identically zero so that we assume that this system leads to $T^\lambda_{\mu\nu} = 0$. Then the cur\-va\-tu\-re tensor $S^\lambda_{\mu \nu\rho}$ is Riemannian one $R^\lambda_{\mu \nu\rho}$. The second term in (\ref{Fad action+Imm term 2}) is identically zero by the properties of the Riemannian tensor and we are left with the purely Einstein action.

Some notable feature of the Faddeev gravity is finiteness of the action even if the field $f^\lambda_A$ is discontinuous. This is connected with that the Faddeev action for gravity does not contain the square of any derivative. The fields with the stepwise coordinate dependence are some particular case of the discontinuous fields, and these can approximate any given field arbitrarily closely. This reminds studying the collection of the flat 4-simplices (4D tetrahedrons) instead of the general Riemannian manifold in the usual general relativity known as Regge calculus \cite{Regge}. In fact, the piecewise flat spacetimes can be described as the simplicial complexes composed of the flat 4-simplices \cite{piecewise flat=simplicial}. In some piecewise affine frame the metric is constant inside the 4-simplices, and analogously the Faddeev fields can be chosen piecewise constant on the general simplicial complex. Unlike the Regge calculus, the 4-simplices in the case of the Faddeev gravity do not necessarily match on their common faces since the tetrad $f^\lambda_A$ and therefore the metric are allowed to be discontinuous.

The aim of this note is just obtaining an analogue of the Regge action by the direct evaluation of the Faddeev action on the piecewise flat spacetime. Our approach is analogous to that one of the work \cite{Fried} where the Regge action has been derived by estimating the Einstein action on the $\delta$-function curvature distribution with support on the triangles. The Faddeev action has been shown to be contributed by the triangles, and the contribution from the triangle has been evaluated for the genuine Faddeev action in our paper \cite{our1}. In the present paper, we take into account the parity odd term in the action and deeper study the structure of the result towards the form of the discrete equations of motion.

\section{Faddeev action on the piecewise constant fields}

Let us write out the Faddeev action (\ref{Fad action+Imm term}) for the piecewise-constant fields on the simplicial complex. Let $x^\lambda$ be a piecewise-affine coordinate frame; $f^\lambda_A (x ) = const$ in the interior of every 4-simplex $\sigma^4$. The field $f^\lambda_A$ in the most part of the neighborhood of any 3-simplex depends (in a stepwise manner) only on the orthogonal to $\sigma^3$ coordinate. Therefore, the contribution to $S$ from $\sigma^3$ is zero. Contributions to $S$ come from the neighborhood of the 2-simplices $\sigma^2$ due to a dependence on the two coordinates, say, $x^1$, $x^2$. Evidently, the expressions $(f^\lambda_{A, \lambda} f^\mu_{B, \mu} - f^\mu_{A, \lambda} f^\lambda_{B, \mu})$ and $\epsilon^{\lambda \mu \nu \rho} f_{\lambda A, \mu} f_{\mu B, \rho}$ appearing in $S$ has support on $\sigma^2$. That is, these are proportional to the $\delta$-function $const \cdot \delta (x^1) \delta (x^2)$. The constants can be reliably defined with taking into account the fact that these expressions are some full derivatives,
\begin{eqnarray}\label{dfdf}                                             
f^\lambda_{A, \lambda} f^\mu_{B, \mu} - f^\mu_{A, \lambda} f^\lambda_{B, \mu} = \partial_\lambda ( f^\lambda_A f^\mu_{B, \mu} - f^\mu_A f^\lambda_{B, \mu}), \\
\label{eps-dfdf} \epsilon^{\lambda \mu \nu \rho} f_{\lambda A, \mu} f_{\mu B, \rho} = \partial_\mu \epsilon^{\lambda \mu \nu \rho} f_{\lambda A} f_{\mu B, \rho}.
\end{eqnarray}

\noindent Then the integral over any neighborhood of the point $(x^1, x^2) = (0, 0)$ (which defines this constant) reduces to the contour integral not depending on the details of the behavior of the fields at this point. Thus we have
\begin{eqnarray}\label{oint_C}                                          
\int (f^\lambda_{A, \lambda} f^\mu_{B, \mu} - f^\mu_{A, \lambda} f^\lambda_{B, \mu}) \d x^1 \d x^2 = \oint_C (f^1_A \d f^2_B - f^2_A \d  f^1_B), \\ \label{oint_C1}
\int \epsilon^{\lambda \mu \nu \rho} f_{\lambda A, \mu} f_{\mu B, \rho} \d x^1 \d x^2 = \oint_C (f_{4 A} \d f_{3 B} - f_{3 A} \d  f_{4 B}).
\end{eqnarray}

\begin{figure}
\unitlength 1pt
\begin{picture}(200,200)(-200,-100)
\put(0,0){\circle{41}}
\put(16,12){\vector(-1,1){4}}
\put(41,82){$2$}
\put(105,-3){$1$}
\put(-49,60){$1^*$}
\put(40,-25){$\sigma^3_{i-1}$}
\put(66,-65){$2^*$}
\put(-44,-10){$\sigma^2=(034)$}
\put(17,15){$C$}
\put(-53,35){\dots}
\put(40,20){$\sigma^4_i$}
\put(-10,40){$\sigma^4_{i+1}$}
\put(10,-50){\dots}

\thicklines

\put(0,0){\line(1,2){40}}
\put(0,0){\line(-2,3){40}}
\put(10,-10){\line(1,-1){50}}
\put(60,-60){\line(2,3){40}}
\put(0,0){\line(1,0){100}}
\put(100,0){\line(-3,4){60}}
\put(-40,60){\line(4,1){80}}

\end{picture}

\caption{To the contribution of $\sigma^2$ to $S$ and the part of $S$ depending on $f^\lambda_A (\sigma^4_i )$.}

\label{sigma2}
\end{figure}
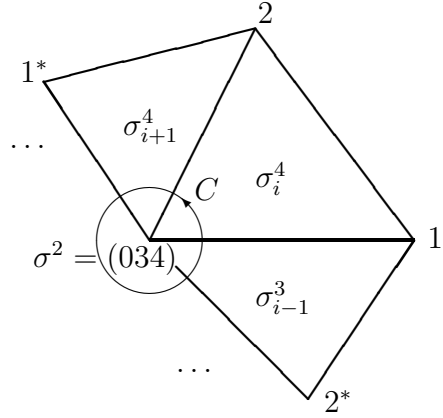

\noindent In fig. \ref{sigma2}, the center $O$ which represents the 2-simplex $\sigma^2$ is encircled by the integration contour $C$ counterclockwise.

Taking into account the subsequent symmetrization over $A, B$ we can permute $A, B$ in the last terms in the RHSs of eqs (\ref{oint_C},\ref{oint_C1}). Thus we have to evaluate the integrals of the type of
\begin{equation}\label{oint_C_general}                                     
\oint_C ( \psi \d \chi - \chi \d \psi )
\end{equation}

\noindent where the functions $\psi , \chi$ of $x^1 , x^2$ are constant almost everywhere and are discontinuous on passing through a few straight lines radiating from some conical singularity on the plane (with a cut beginning at the singularity). An ambiguity may be connected with defining the products of the step functions and delta functions arising under the contour integral sign. Let $\sigma^4_1$, \dots, $\sigma^4_{i-1}$, $\sigma^4_{i}$, $\sigma^4_{i+1}$, \dots, $\sigma^4_n$ be the 4-simplices encountered in moving along the contour $C$ and sharing the 2-simplex $\sigma^2$. If the point $( x^1 , x^2 )$ is located on (the 2D projection of) the 3-face $\sigma^4_{i} \cap \sigma^4_{i+1}$ separating the 4-simplices $\sigma^4_{i}$ and $\sigma^4_{i+1}$, the values of the functions $\psi , \chi$ are ambiguous; let us parameterize these by some $\alpha$,
\begin{equation}                                                           
\psi ( \sigma^4_{i} \cap \sigma^4_{i+1} ) = (1 - \alpha ) \psi (\sigma^4_i ) + \alpha \psi (\sigma^4_{i+1} ), ~~~ \chi ( \sigma^4_{i} \cap \sigma^4_{i+1} ) = (1 - \alpha ) \chi (\sigma^4_i ) + \alpha \chi (\sigma^4_{i+1} ).
\end{equation}

\noindent Here, we have naturally proposed a symmetry between $f^1_A$ and $f^2_A$, $f_{4 A}$ and $f_{3 A}$, and have taken the same $\alpha$ for both $\psi$ and $\chi$. Then the value of the integral (\ref{oint_C_general}) can be readily found,
\begin{eqnarray}\label{delta f delta f_general}                            
\oint_C ( \psi \d \chi - \chi \d \psi ) & = & \sum^n_{i = 1} \left \{ [ \alpha \psi (\sigma^4_{i+1} ) + (1 - \alpha ) \psi (\sigma^4_i ) ] [ \chi (\sigma^4_{i+1} ) - \chi (\sigma^4_i ) ] \right. \nonumber \\ & & \left. - [ \alpha \chi (\sigma^4_{i+1} ) + (1 - \alpha ) \chi (\sigma^4_i ) ] [ \psi (\sigma^4_{i+1} ) - \psi (\sigma^4_i ) ] \right \} \nonumber \\ & = & \sum^n_{i=1} [ \psi ( \sigma^4_i ) \chi ( \sigma^4_{i+1} ) - \psi ( \sigma^4_{i+1} ) \chi ( \sigma^4_i ) ].
\end{eqnarray}

\noindent Remarkable is that the dependence on $\alpha$ has disappeared. It is easy to see that (\ref{delta f delta f_general}) is zero if either $\phi (\sigma^4_i )$ or $\chi (\sigma^4_i )$ does not depend on $\sigma^4_i$. This means that (\ref{delta f delta f_general}) depends only on the variations of $\psi$ and $\chi$ from 4-simplex to 4-simplex. Now we can define the constants at the delta-functions,
\begin{eqnarray}\label{dfdf,epsilondfdf}                                
f^\lambda_{A, \lambda} f^\mu_{B, \mu} - f^\mu_{A, \lambda} f^\lambda_{B, \mu} = \delta (x^1 ) \delta (x^2 ) \sum^n_{i=1} [ f^1_A ( \sigma^4_i ) f^2_B ( \sigma^4_{i+1} ) - f^1_A ( \sigma^4_{i+1} ) f^2_B ( \sigma^4_i ) ], \\ \epsilon^{\lambda \mu \nu \rho} f_{\lambda A, \mu} f_{\mu B, \rho} = \delta (x^1 ) \delta (x^2 ) \sum^n_{i=1} [ f_{4 A} ( \sigma^4_i ) f_{3 B} ( \sigma^4_{i+1} ) - f_{4 A} ( \sigma^4_{i+1} ) f_{3 B} ( \sigma^4_i ) ].
\end{eqnarray}

\noindent The symmetrization over $A, B$ is implied.

We substitute these expressions into (\ref{Fad action+Imm term}). Then we have the products of $\delta (x^1 ) \delta (x^2 )$ and $\Pi^{AB} \sqrt{g}$ or $\Pi^{AB}$ discontinuous at $(x^1, x^2) \to (0, 0)$. These products are defined ambiguously depending on the intermediate regularization of the discontinuities. We can absorb this ambiguity into some effective values $\Pi^{AB} \sqrt{g}$ or $\Pi^{AB}$ on the 2-simplex $\sigma^2$, $\Pi^{AB} (\sigma^2 ) \sqrt{g (\sigma^2 )}$ and $\Pi^{AB} (\sigma^2 )$, and write
\begin{eqnarray}\label{dfdfVPi+epsilondfdfPi}                           
(f^\lambda_{A, \lambda} f^\mu_{B, \mu} - f^\lambda_{A, \mu} f^\mu_{B, \lambda}) \Pi^{AB} \sqrt{g} & = & \delta (x^1 ) \delta (x^2 ) \Pi^{AB} (\sigma^2 ) \sqrt{g (\sigma^2 )} \sum^n_{i=1} \left [ f^1_A ( \sigma^4_i ) f^2_B ( \sigma^4_{i+1} ) \right. \nonumber \\ & & \left. - f^1_A ( \sigma^4_{i+1} ) f^2_B ( \sigma^4_i ) \right ], \\
\label{epsilondfdfPi}
\epsilon^{\lambda \mu \nu \rho} f_{\lambda A, \mu} f_{\mu B, \rho} \Pi^{AB} & = & \delta (x^1 ) \delta (x^2 ) \Pi^{AB} (\sigma^2 )  \sum^n_{i=1} \left [ f_{4 A} ( \sigma^4_i ) f_{3 B} ( \sigma^4_{i+1} ) \right. \nonumber \\ & & \left. - f_{4 A} ( \sigma^4_{i+1} ) f_{3 B} ( \sigma^4_i ) \right ].
\end{eqnarray}

\noindent

To write out the action in an invariant form, it is convenient to decompose the world vectors into the components along some four vectors of edges,
\begin{equation}\label{contravar-edge}                                     
f^\lambda_A (\sigma^4_i ) = \sum_\mu f^{\sigma^1_\mu}_A (\sigma^4_i ) \Delta x^\lambda_{\sigma^1_\mu}.
\end{equation}

\noindent Here, the four edges $\sigma^1_\mu$, $\mu = 1, 2, 3, 4$, span some 4-simplex $\sigma^4_i$. Besides that,
\begin{equation}                                                           
\Delta x^\lambda_{\sigma^1} = x^\lambda (\sigma^0_2 ) - x^\lambda (\sigma^0_1 )
\end{equation}

\noindent for the edge $\sigma^1$, the difference between the coordinates of its ending vertices $\sigma^0_1$, $\sigma^0_2$. The "contravariant edge components" $f^{\sigma^1_\lambda}_A$ are world invariants.

Let us substitute this into equation (\ref{dfdfVPi+epsilondfdfPi}). The $\delta$-function factor in (\ref{dfdfVPi+epsilondfdfPi}) being integrated over $\d^4 x$ leads to an integral over $\d x^3 \d x^4$ over $\sigma^2$ which is equal to
\begin{equation}                                                           
\frac{1}{2} \left ( \Delta x^3_{\sigma^1_3} \Delta x^4_{\sigma^1_4} - \Delta x^3_{\sigma^1_4} \Delta x^4_{\sigma^1_3} \right )
\end{equation}

\noindent where the edges $\sigma^1_3$, $\sigma^1_4$ span the triangle $\sigma^2$. The integral of (\ref{dfdfVPi+epsilondfdfPi}) over $\d^4 x$ gives a contribution of $\sigma^2$ into the main (without the parity-odd term) part of the action,
\begin{eqnarray}\label{df df Dx Dx Dx Dx V Pi}                             
S_{\rm main} (\sigma^2 ) & = & \frac{1}{4} \Pi^{AB} (\sigma^2 ) \sqrt{g (\sigma^2 )} \sum^n_{i=1} \left [ f^{\tson_1}_A ( \sigma^4_i ) f^{\tson_2}_B ( \sigma^4_{i+1} ) - f^{\tson_1}_A ( \sigma^4_{i+1} ) f^{\tson_2}_B ( \sigma^4_i ) \right ] \nonumber \\ & & \cdot \left ( \Delta x^1_{\tson_1} \Delta x^2_{\tson_2} - \Delta x^1_{\tson_2} \Delta x^2_{\tson_1} \right ) \left ( \Delta x^3_{\son_3} \Delta x^4_{\son_4} - \Delta x^3_{\son_4} \Delta x^4_{\son_3} \right ).
\end{eqnarray}

\noindent Here, $\tson_1$, $\tson_2$ are dummy indices running over $\son_1$, $\son_2$, $\son_3$, $\son_4$, the summation over them is implied. The above independence from $x^3$, $x^4$ in our construction in this section assumes that $\Delta x^1_{\son_3} = \Delta x^2_{\son_3} = \Delta x^1_{\son_4} = \Delta x^2_{\son_4} = 0$ (that is, $\sigma^2$ is located in the $(x^3, x^4)$-plane). Then the action (\ref{df df Dx Dx Dx Dx V Pi}) reads
\begin{equation}\label{df df Pi V det Dx}                                  
\hspace{-5mm} S_{\rm main} (\sigma^2 ) = \frac{1}{2} \Pi^{AB} (\sigma^2 ) \sqrt{g (\sigma^2 )} \det \| \Delta x^\lambda_{\son_\mu} \| \sum^n_{i=1} \left [ f^{\son_1}_A ( \sigma^4_i ) f^{\son_2}_B ( \sigma^4_{i+1} ) - f^{\son_1}_A ( \sigma^4_{i+1} ) f^{\son_2}_B ( \sigma^4_i ) \right ].
\end{equation}

\noindent Here, $\sqrt{g (\sigma^2 )} \det \| \Delta x^\lambda_{\son_\mu} \|$ is a world invariant (a volume), and it can be written as simply $\sqrt{g (\sigma^2 )}$ if the metric tensor forming this determinant is implied to be taken in the components projected on the edges, $g_{\sigma^1_1 \sigma^1_2} = g_{\lambda \mu} \Delta x^\lambda_{\sigma^1_1 } \Delta x^\mu_{\sigma^1_2 }$. The action (\ref{df df Pi V det Dx}) itself is explicitly a world invariant too.

Integrating (\ref{epsilondfdfPi}) and recasting the result in an invariant form is performed even easier using the world invariants obtained by projecting the covariant tetrad components along the edges,
\begin{equation}\label{f-sigma-mu}                                         
f^A_{\sigma^1_\mu} = f^A_\lambda \Delta x^\lambda_{\sigma^1_\mu}.
\end{equation}

\noindent The contribution of $\sigma^2$ to the parity odd part of the action (\ref{Fad action+Imm term}) takes the form
\begin{equation}\label{int-epsilon-dfdfPi}                                 
\hspace{-15mm} S_{\rm odd}(\sigma^2) = - \frac{1}{2 \gamma_F} \Pi^{AB} (\sigma^2 ) \sum^n_{i=1} \left [ f_{\son_4 A }( \sigma^4_i ) f_{\son_3 B} ( \sigma^4_{i+1} ) - f_{\son_4 A } ( \sigma^4_{i+1} ) f_{\son_3 B} ( \sigma^4_i ) \right ].
\end{equation}

\noindent The sums in both the (\ref{df df Pi V det Dx}) and (\ref{int-epsilon-dfdfPi}) are, in fact, bilinears in the variations of the fields $f$ from 4-simplex to 4-simplex.

Now let us concentrate on the terms in the action containing the fields in the above given 4-simplex $\sigma^4_i$. To this end, we denote its vertices by 0, 1, 2, 3, 4, see fig. \ref{sigma2}. Let $k, l, m, \dots$ run over 0, 1, 2, 3, 4 (vertices), $(kl), (klm), (klmn)$ and $(klmnp) = (01234)$ denote the 1-, 2-, 3- and 4-simplices, respectively, with the vertices listed in the parentheses (unordered sets). The contribution of the 2-simplices belonging to $\sigma^4_i$ are of interest. To evaluate these, we need to expand $f^\lambda_A$ according to (\ref{contravar-edge}) over $\Delta x^\lambda_{\sigma^1_\mu}$ for the different tetrads $\sigma^1_\mu$ of the edges in $\sigma^4_i$. Consider re-expanding a world vector $A^\lambda$ when passing from the edges emanating from the vertex 0 to the edges emanating from 4:
\begin{eqnarray}                                                           
A^\lambda & = & A^{01} \Delta x^\lambda_{01} + A^{02} \Delta x^\lambda_{02} + A^{03} \Delta x^\lambda_{03} + A^{04} \Delta x^\lambda_{04}, \nonumber \\ A^\lambda & = & A^{40} \Delta x^\lambda_{40} + A^{41} \Delta x^\lambda_{41} + A^{42} \Delta x^\lambda_{42} + A^{43} \Delta x^\lambda_{43}, ~~~ \Delta x^\lambda_{kl} = x^\lambda_l - x^\lambda_k.
\end{eqnarray}

\noindent Here, the {\it ordered} pairs $kl$ denote the {\it oriented} edges. Then
\begin{equation}                                                           
A^{40} = - A^{01} - A^{02} - A^{03} - A^{04}, ~~~ A^{41} = A^{01}, ~~~ A^{42} = A^{02}, ~~~ A^{43} = A^{03}.
\end{equation}

\noindent It is seen that the contravariant edge components should be associated with the vertices of the 4-simplex rather than four edges, and these are, of course, not independent,
\begin{equation}                                                           
A^{kl} \equiv A^l, ~~~ \sum^4_{k = 0} A^k = 0.
\end{equation}

As for the covariant world components, these being projected on the ten edges form quite a redundant set constrained, however, by simple closure conditions,
\begin{equation}                                                           
A_{kl} = A_\lambda \Delta x^\lambda_{kl}, ~~~ A_{kl} + A_{lm} + A_{mk} = 0.
\end{equation}

The terms in the action depending on $f^\lambda_A (\sigma^4_i)$ also depend on the fields in the 4-simplices sharing with $\sigma^4_i$ its 3-faces $(lmnp)$; these are denoted by $(k^*lmnp)$ (up to any permutation of the symbols in the parentheses).

The terms in the action depending on $f^\lambda_A (\sigma^4_i) \equiv f^\lambda_A (01234)$ take the form
\begin{eqnarray}\label{S-sigma4}                                           
S(\sigma^4_i) = \frac{1}{2} \sum_{(kl)} \Pi^{AB} (mnp) \sqrt{g (mnp)} \left \{ f^k_A (01234) \left [ f^l_B (k^*lmnp) - f^l_B (kl^*mnp) \right ] \right. \nonumber \\ \left. - f^l_B (01234) \left [ f^k_A (k^*lmnp) - f^k_A (kl^*mnp) \right ] \right \} \nonumber \\ - \frac{1}{2 \gamma_F} \sum_{(kl)} \Pi_{AB} (mnp) \left \{ f^A_{mn} (01234) \left [ f^B_{mp} (k^*lmnp) - f^B_{mp} (kl^*mnp) \right ] \right. \nonumber \\ \left. - f^B_{mp} (01234) \left [ f^A_{mn} (k^*lmnp) - f^A_{mn} (kl^*mnp) \right ] \right \}.
\end{eqnarray}

\noindent Here the summation is over the ten 2-simplices $(mnp)$ of the 4-simplex $(01234)$ or the ten edges $(kl)$ dual to $(mnp)$ and
\begin{equation}\label{closing-f}                                          
\sum^4_{k=0} f^k_A = 0, ~~~ f^A_{kl} + f^A_{lm} + f^A_{mk} = 0.
\end{equation}

\section{Conclusion}

Due to the relations (\ref{closing-f}), the sum of the terms in $S(\sigma^4_i)$ referred to the same pair of the 4-simplices, say, $(01234)$ and $(k^*lmnp)$ is zero if $\Pi^{AB} (mnp) \sqrt{g (mnp)}$ and $\Pi^{AB} (mnp)$ do not depend on $(mnp)$. On discrete level, this reflects the fact that the multipliers of $\Pi^{AB} \sqrt{g }$ and $\Pi^{AB} $ in the continuum theory are full derivatives, see (\ref{dfdf}), (\ref{eps-dfdf}). Also this means that the action depends on the differences of $\Pi^{AB} (mnp) \sqrt{g (mnp)}$ and $\Pi^{AB} (mnp)$ for the different 2-simplices $(mnp)$ rather than on these factors themselves. Taking into account this fact and also that $S( \sigma^2 )$ is bilinear in the variations of the fields $f$ from 4-simplex to 4-simplex, we can conclude that both the action itself and the equations of motion obtained by differentiating (\ref{S-sigma4}) with respect to $f^\lambda_A (01234)$ are bilinear in the variations of the fields $f$ from 4-simplex to 4-simplex. This is consistent, in particular, with that the discrete vertical equations have the continuum counterpart (\ref{V lambda A-modified}) bilinear in the derivatives. So we can expect that if the given discrete variables $f$ describe an approximation to some fixed smooth manifold (continuum distribution of $f$), then making this discretization finer and finer we get the discrete vertical equations approximating the continuum counterpart with the accuracy $O(\varepsilon )$ where $\varepsilon$ is some typical edge length. In this procedure, the effective values $\Pi^{AB} (\sigma^2 ) \sqrt{g (\sigma^2 )}$ and $\Pi^{AB} (\sigma^2 )$ can be defined by taking the values of the continuum field $f$ distribution in the neighborhood of $\sigma^2$.

An interesting problem is whether the definition of $\Pi^{AB} (\sigma^2 ) \sqrt{g (\sigma^2 )}$, $\Pi^{AB} (\sigma^2 )$ (in terms of the values of the fields $f$ in the 4-simplices) exists such that the discrete vertical equations would lead {\it exactly} to the discrete general relativity (Regge calculus).

The author thanks I.A. Taimanov who had attracted author's attention to the new formulation of gravity and Ya.V. Bazaikin for valuable discussions of this subject. The author is grateful to I.B. Khriplovich who has provided moral support, A.A. Pomeransky and A.S.Rudenko for discussion at a seminar, stimulating the writing of this article. The present work was supported by the Ministry of Education and Science of the Russian Federation, Russian Foundation for Basic Research through Grants No. 09-01-00142-a, 11-02-00792-a and Grant 14.740.11.0082 of federal program "personnel of innovational Russia".


\end{document}